# System for Analysis of Wind Collocations (SAWC): A Novel Archive and Collocation Software Application for the Intercomparison of Winds from Multiple Observing Platforms

Katherine E. Lukens [1,2,*], Kevin Garrett [3], Kayo Ide [4], David Santek [5], Brett Hoover [6], David Huber [7], Ross N. Hoffman [1,2], and Hui Liu [1,2]

[1] NOAA/NESDIS/Center for Satellite Applications and Research (STAR), College Park, Maryland 20740 USA
[2] Cooperative Institute for Satellite Earth System Studies (CISESS), University of Maryland, College Park, Maryland 20740 USA
[3] NOAA/NWS/Office of Science and Technology Integration (OSTI), Silver Spring, Maryland 20910 USA
[4] University of Maryland, College Park, Maryland 20740 USA
[5] Cooperative Institute for Meteorological Satellite Studies (CIMSS), University of Wisconsin-Madison, Madison, WI 53706 USA
[6] Lynker Technologies at NOAA/NWS/NCEP/Environmental Modeling Center (EMC), College Park, Maryland 20740 USA
[7] Redline Performance Solutions, LLC at NOAA/NWS/NCEP/Environmental Modeling Center (EMC), College Park, Maryland 20740 USA
[*] Correspondence: katherine.lukens@noaa.gov

**Abstract:** Accurate atmospheric 3D wind observations are a high priority in the science community. To address this requirement and to support researchers' needs to acquire and analyze wind data from multiple sources, the System for Analysis of Wind Collocations (SAWC) was jointly developed by NOAA/NESDIS/STAR, UMD/ESSIC/CISESS, and UW-Madison/CIMSS. SAWC encompasses a multi-year archive of global 3D winds observed by Aeolus, sondes, aircraft, stratospheric superpressure balloons, and satellite-derived atmospheric motion vectors, archived and uniformly formatted in netCDF for public consumption; identified pairings between select datasets collocated in space and time; and a downloadable software application developed for users to interactively collocate and statistically compare wind observations based on their research needs. The utility of SAWC is demonstrated by conducting a one-year (September 2019-August 2020) evaluation of Aeolus level-2B (L2B) winds (Baseline 11 L2B processor version). Observations from four archived conventional wind datasets are collocated with Aeolus. Recommended quality controls are applied. Wind comparisons are assessed using the SAWC collocation application. Comparison statistics are stratified by season, geographic region, and Aeolus observing mode. The results highlight the value of SAWC's capabilities, from product validation through intercomparison studies to the evaluation of data usage in applications and advances in the global Earth observing architecture.

## 1. Introduction

In 2018, the National Academies Press released the 2017-2027 decadal survey which listed accurate 3D wind observations as a top priority in the Earth science community (National Academies of Sciences, Engineering, and Medicine, 2018), where 3D indicates the observed winds are spread over all 3 geometric dimensions although the observations are actually 2-dimensional (2D) horizontal winds. Wind observations have a critical impact on weather forecasting, data assimilation (DA), ocean currents, wildfires, air quality, the energy sector, the spread of disease, and other applications. For decades, winds have been observed in situ by commercial aircraft, sondes (e.g., rawin- and dropsondes), and near-space superpressure balloons (Hertzog et al., 2006; Morel & Bandeen, 1973; Rabier et al., 2010; Rhodes & Candido, 2021). Winds called atmospheric motion vectors (AMVs) are also derived by tracking moisture and water vapor features in satellite imagery through time (e.g., Velden et al., 1997; Santek et al., 2014, 2019; Cotton et al., 2020). Additionally, Doppler wind lidar onboard the Aeolus satellite has successfully provided wind information on a global scale (Reitebuch et al., 2009; Stoffelen et al., 2005).

Numerous studies have been conducted to address the decadal survey's 3D wind requirement (e.g., Velden and Holmlund, 1998; Bedka et al., 2009; Velden and Bedka, 2009; Bormann et al., 2003; Genkova et al., 2008, 2010; Santek et al., 2014, 2019, 2021, 2022; Rani et al., 2022; Borde et al., 2016, 2019; Martin et al., 2021; Hoffman et al., 2022; Lukens et al., 2022, 2023a). Many involve the characterization and validation of AMVs, e.g., from establishing collocation standards between sondes and AMVs (time difference < 90 minutes, height difference < 25 hPa, horizontal distance <

150 km) (Velden and Holmlund, 1998) to investigating the importance of height assignment accuracy to and spatial correlations of AMV observation errors for DA (Velden and Bedka, 2009; Bormann et al., 2003), as well as intercomparing AMVs derived by several international satellite wind producing centers (Genkova et al., 2008; Santek et al., 2014, 2019, 2022). Recent studies now include Aeolus in their comparisons, e.g, to validate Aeolus winds (Martin et al., 2021; Rani et al., 2022), and to leverage Aeolus as a potential comparison standard for AMV characterization (Hoffman et al., 2022; Lukens et al., 2022).

Results from these different studies are not easily comparable, due in part to each study using widely varying collocation criteria. As there are few accepted collocation standards for different wind pairings in the literature and broader wind community, researchers often find the need to create their own criteria and algorithms. The standards that do exist were developed decades ago yet are still used today, i.e., those for sondes versus AMV comparisons (Velden and Holmlund, 1998). It has been recommended that such standards be reassessed to account for today's advanced imagers (Daniels, 2022). Further, it is often necessary for researchers to acquire and reformat disparate data from numerous sources, and develop their own analysis tools in order to achieve specific project objectives. These steps can be time-consuming and delay progress, particularly for projects assigned short periods of performance.

To support researchers' needs to acquire and analyze wind data from multiple sources, and to further address the 3D wind requirement, the System for Analysis of Wind Collocations (SAWC) was jointly developed by the NOAA/NESDIS/Center for Satellite Applications and Research (STAR), the Cooperative Institute for Satellite Earth System Studies (CISESS) at the University of Maryland, and the Cooperative Institute for Meteorological Satellite Studies (CIMSS) at the University of Wisconsin-Madison. SAWC is unique in that it provides the data and tools one needs for wind observation research all in one place: a multi-year, public archive of global 3D winds that have been acquired from multiple observing platforms and converted to a common format (netCDF); identified pairings between select datasets also available in netCDF; and a downloadable software application developed for users to interactively collocate wind observations and visually and statistically compare them based on their research needs. The capability of SAWC is wide-ranging, from product validation through intercomparison studies to the evaluation of data usage in applications and advancements in the global Earth observing architecture.

This article introduces SAWC and its components, including the collocation application and a multi-year collection of wind observations from Aeolus, satellite imagery (AMVs), aircraft, sondes, and Loon stratospheric superpressure balloons (Rhodes and Candido, 2021). Examples of SAWC's utility are presented, including a one-year quantitative evaluation of Aeolus winds compared to four conventional wind datasets and an example assessment of observation error estimates in DA.

Section 2 provides an overview of SAWC and its components. Section 3 discusses example applications of SAWC. Section 4 presents a summary and conclusions.

**2. Overview of SAWC**

SAWC encompasses a global wind data archive (see Section 2.1), identified pairings between select datasets in the form of index files (see Section 2.2.1), a collocation software application consisting of a collocation tool (see Section 2.2.1) and a plotting tool (see Section 2.2.2), and a user manual (Lukens et al., 2023b). All components of SAWC are hosted and maintained on the NOAA/NESDIS/STAR public server, are continuously updated, and are publicly available online:

- SAWC Website: https://www.star.nesdis.noaa.gov/sawc
- Data Home: https://www.star.nesdis.noaa.gov/data/sawc
- Wind Archive: https://www.star.nesdis.noaa.gov/data/sawc/wind_datasets
- Software Application & Index Files: https://www.star.nesdis.noaa.gov/data/sawc/collocation
- User Manual: https://www.star.nesdis.noaa.gov/data/sawc/User_Manual

SAWC's end-to-end process comprises three main steps: (1) Data Acquisition, (2) Collocation of Winds, and (3) Analysis and Visualization (Fig. 1). In Step (1), data is acquired from various sources, reformatted to a common format (netCDF), and archived for public dissemination (hereafter source data files). Note that the user does not interface with Step (1); rather, the data files are archived in netCDF and then are made available to the user.

The user interface with SAWC occurs during Steps (2) and (3). In Step (2), the application's collocation tool is used to ingest and match winds from user-selected source data and generate index files, i.e., netCDF files that contain the array indices or locations of the matched winds in the source data files. In Step (3), the application's plotting tool uses these index files to extract and compare the matched winds through visual and statistical analysis. A host of figures are generated that quantify the comparisons. See Section 2.2 for more on the collocation application.

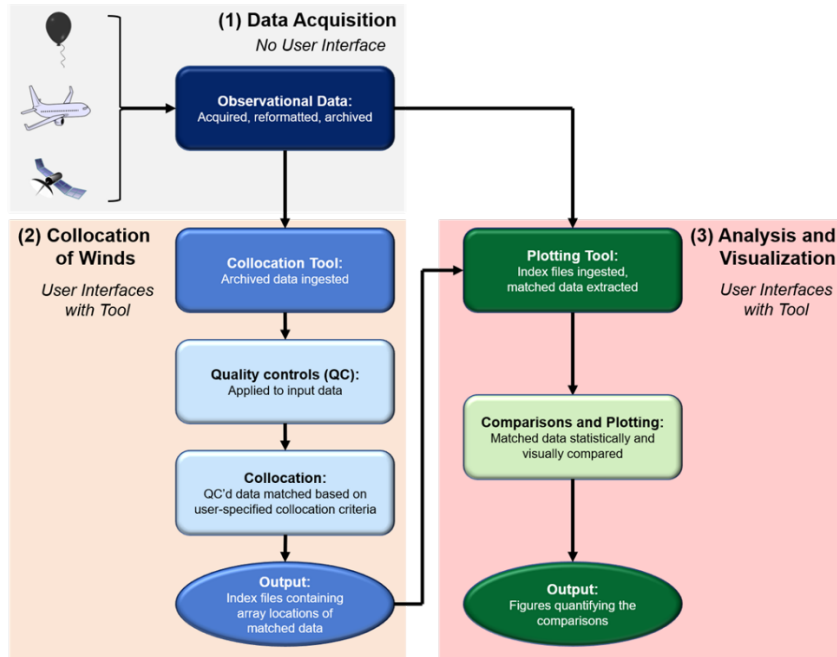

**Figure 1.** Flow chart illustrating SAWC's end-to-end process. Note that the user interfaces with the collocation and plotting tools in Steps (2) and (3), respectively.

*2.1. Datasets Available*

At the time of writing, several global wind observation datasets have been acquired and archived. These include Doppler wind lidar (DWL) observations from Aeolus, satellite-derived winds (i.e., AMVs), and in situ upper-level winds observed by sondes, stratospheric superpressure balloons, and aircraft. These datasets are included in the SAWC archive as a first step; certainly, SAWC can be expanded to include others such as surface, radial, and profiler winds which have yet to be included as they are out of the scope of the wind comparison study presented in Section 3. Table l lists the file formats and temporal coverage of the archived winds currently available in SAWC, as well as a few key variables provided in each dataset.

**Table 1.** Source wind datasets currently available in SAWC. The file formats, temporal coverage, vertical coordinates, and wind representation variables available per dataset are listed. Additionally, all datasets include time variables (year, month, day, and hour) as well as horizontal coordinates (latitude and longitude).

| Wind Datasets | SAWC File Formats | Temporal Coverage | Vertical Coordinates | Wind Representation |
|---|---|---|---|---|
| Aeolus | NetCDF; BUFR; ESA's Earth Explorer (EE) format | Sep 2018 - Apr 2023 | Height; Pressure | HLOS Wind Velocity; Azimuth Angle |
| Loon | NetCDF-4 | 2011 - 2021 | Height; Pressure | u-/v-components; Wind Direction |
| Sonde | NetCDF-4 | Sep 2018 - Present Day | Height; Pressure | Wind Speed; Wind Direction |
| Aircraft | NetCDF-4 | Sep 2018 - Present Day | Height | Wind Speed; Wind Direction |
| AMV | NetCDF-4 | Sep 2018 - Present Day | Pressure | Wind Speed; Wind Direction |

Aeolus level-2B (L2B) winds were provided by the European Space Agency (ESA) and the European Centre for Medium-Range Weather Forecasts (ECMWF). Aeolus winds were observed by the DWL onboard the Aeolus satellite that traveled on a dawn/dusk orbit during 2018-2023. The Aeolus lidar was aimed to the right of the spacecraft so that horizontal line-of-sight (HLOS) winds were observed in a plane that intersects the Earth's surface 230 km to the right of the sub-satellite track with an incidence angle of about 37.6° (Reitebuch et al., 2018). Two main wind regimes were retrieved for use in research and operations: Rayleigh-clear (RayClear) winds were derived from molecular backscattering and represented winds in clear scenes, and Mie-cloudy (MieCloud) winds were derived from aerosol backscattering and represented winds in cloudy scenes. Note that the data quality changes over the course of Aeolus' lifetime due to the DWL signal loss (Straume et al., 2021) and because the L2B processor was upgraded several times. The expected spatial coverage of Aeolus for a 24-h period is displayed in Fig. 2a.

AMV, aircraft, and sonde winds are provided by NCEP and are processed by the NCEP/EMC Obsproc team prior to their conversion to netCDF and archival in SAWC. Aircraft winds are observed at flight level in the upper troposphere/lower stratosphere and in the ascending and descending legs of each flight. Spatial coverage is regional and dependent upon commercial flight paths. An example of the daily coverage of aircraft observations is shown in Fig. 2b. All available AMVs are archived and are from both geostationary (GEO) and polar Earth-orbiting (LEO) satellites. Spatial coverage for the entire dataset is near-global at various vertical levels (Fig. 2c). Sondes take direct in situ wind observations as they ascend through the atmosphere from the surface to the upper troposphere/lower stratosphere. Spatial coverage is sparse and mostly over land (Fig. 2d). Note that all SAWC-archived winds from aircraft, satellite imagery, and sondes contain only unrestricted observations.

Stratospheric superpressure balloon observations were provided by Loon, a former subsidiary of Google's parent company Alphabet. From 2011 to 2021, Loon deployed a network of these tennis court-sized balloons each carrying an instrument payload platform used to provide internet connectivity to regions with limited access. The balloon configurations were specifically designed to withstand harsh stratospheric conditions for months at a time. A ground team was able to control their movements by remotely adjusting the balloon pressure, allowing the configurations to move into different airstreams. In situ atmospheric observations were reported at balloon level (~50-100 hPa or 18-20 km) at a frequency of ~1-20 minutes. The spatial coverage was regional (Fig. 2e) and depended upon the Loon mission at the time of deployment.

All datasets span the entire lifetime of Aeolus (Sept 2018 - April 2023), except for the acquisition of AMVs, aircraft, and sonde winds that continues beyond Aeolus' lifetime to Present Day, and the Loon record that runs from 2011 to 2021 when the Loon project ended (see Table 1). All netCDF files contain 6 hours of data centered around four numerical weather prediction (NWP) analysis times (00, 06, 12, and 18 UTC), except for Aeolus netCDF files which cover 24-h periods (00:00-23:59 UTC) with separate files for each Aeolus wind regime. The archived Aeolus data are also available in ECMWF BUFR and ESA's Earth Explorer (EE) formats.

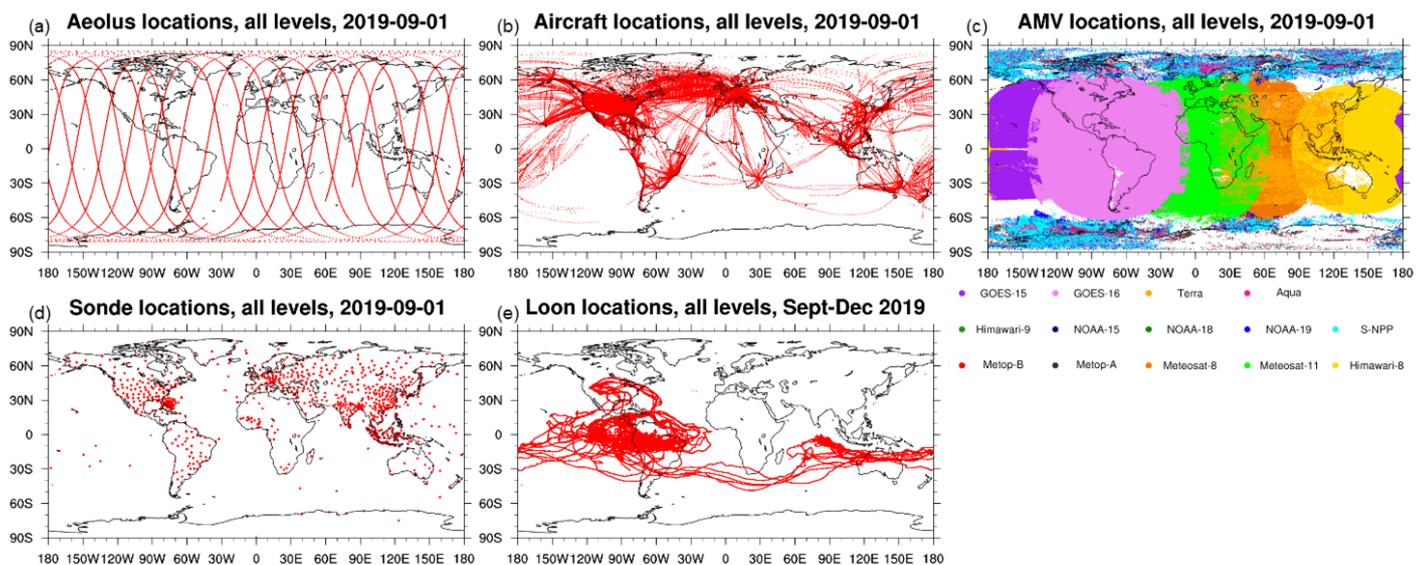

**Figure 2.** Example spatial coverage of each wind dataset currently available in the SAWC archive for a 24-h period for (**a**) Aeolus, (**b**) aircraft, (**c**) AMVs, and (**d**) sondes, and for a 4-month period for (**e**) Loon. Note that in (**c**) each satellite is stratified by color.

*2.2. Collocation Software Application*

Unique collocation techniques were developed at CIMSS (Santek et al., 2021) and incorporated into the SAWC collocation application, a Python-based software package developed to match winds from different datasets based on user-specified criteria. There are two main tools that make up the collocation application: a collocation tool, and a plotting tool. The collocation tool matches wind observations from different datasets and only considers the time and 3D locations of the wind observations, not the wind speed or direction. The plotting tool intercompares the matched wind observations, performs statistical analyses, and generates plots quantifying the comparisons. Table 2 lists the SAWC management components discussed in this section.

**Table 2.** List of SAWC management components: index files and the collocation software application.

| SAWC Component | Formats | Temporal Coverage | Key Variables Available |
|---|---|---|---|
| Index Files | NetCDF-4 files | Select periods | Indices of matched winds; Differences in time, height/pressure, distance for each pair of matched winds |
| Collocation Application | One tarball per application version, each containing Bash and Python scripts | N/A | N/A |

2.2.1. Collocation Tool

To collocate the winds, the user modifies various input parameters for the collocation tool. Table 3 lists the main input parameters the user can change; a complete list is provided in the SAWC user manual (Lukens et al., 2023b).

**Table 3.** List of main input parameters the user can modify in the collocation tool.

| Collocation Tool Parameter | Description |
|---|---|
| Date Range | Year, month, and range of days over which to run the collocation tool. |
| Dataset Names | Names of datasets to be collocated. (The first dataset listed is the Driver; all others are Dependents.) |
| Path to Output Index Files | Full path to location where output collocation index files are to be saved. |
| Collocation Criteria | 4 criteria: Max collocation distance in km; Max time difference in minutes; Max log10(pressure) difference log10(hPa); Max height difference in km. (Must have 4 criteria per Dependent dataset to be collocated.) |
| Quality Control Flags | Quality control (QC) flags for each dataset (Driver and Dependents) indicating whether or not QC will be applied. Options: 0 (no QC applied); 1 (QC applied) |
| Number of Matches Allowed | Number of Dependent observations allowed to match each Driver observation. Default = 50 |
| AMV Quality Indicator | Quality indicator (QI) value in % for AMV observations. |
| AMV Quality Indicator Option | QI option for AMVs. Options: NO_FC (default; use QI variable without forecast); YES_FC (use QI variable with forecast) |
| Aeolus Dataset Type | Abbreviation for Aeolus L2B dataset type. Options: orig (dataset processed with original L2B processor at time of retrieval); B## (dataset reprocessed with a different L2B processor than that used at time of retrieval, where ## is a 2-digit number indicating the Baseline number, e.g., B10 = Baseline 10) |

Prior to collocation, quality controls (QC) are applied to the input wind observations if the user chooses this option. At present, QCs are readily available for two datasets, Aeolus and AMVs, based on recommendations from the data producers and broader wind community (Table 4). It is important to note that the user is able to change the QC parameters and/or add QCs for the other datasets if desired. After QC, a 4D collocation (latitude, longitude, height/pressure, and time) is performed between the "Driver" dataset and one or more "Dependent" datasets, where the Driver is the dataset with which all Dependent datasets are collocated. Four collocation criteria are needed per Dependent dataset: maximum time difference in minutes, maximum collocation distance in km, maximum $\log_{10}$(pressure) difference in $\log_{10}$(hPa), and maximum height difference in km. Users are able to choose their own collocation criteria, if desired, or they can opt to use the default settings (Table 5).

During the collocation process, all Dependent observations are compared to each Driver observation. This means that one Dependent observation may be collocated with multiple Driver observations, or one Driver observation may be collocated with multiple Dependent observations. If a Dependent observation meets all selected collocation criteria, the Driver and Dependent array locations (i.e., indices) in the respective source data files are saved to the output index file for that Driver-Dependent pairing. Index files for select pairings and time periods are available in the SAWC archive. By saving the matched indices and not a copy of the collocated data itself, disk space is saved. In addition, each index file contains the actual distances and differences in time and height/pressure between each pair of collocated winds. The index files can be subsetted with stricter but not looser collocation criteria.

**Table 4.** Wind community-recommended QC parameters currently available in SAWC that are applied to the listed datasets only if the user selects this option. Default reject values are listed in the table. For Aeolus, p is pressure in hPa, sigma is the L2B uncertainty in m/s, z is the height of the vertical range bins within which measurements are accumulated, and length is the integration length in km over which the measurements are accumulated. For AMVs, the quality indicator (QI) is the minimal acceptable percent confidence value (in %).

| Dataset | QC Parameters |
|---|---|
| Aeolus Mie-cloudy | p > 800 hPa; $\sigma$ > 5 m/s |
| Aeolus Rayleigh-clear | p > 800 hPa; $\sigma$ > 8.5 m/s for 800 ≥ p > 200 hPa; $\sigma$ > 12 m/s for p 200 hPa; z < 0.3 km; length < 60 km |
| AMV | QI < 80 |

**Table 5.** Default collocation criteria applied to each listed dataset. t in minutes indicates the maximum time difference allowed between the Driver and Dependent observations; x in km is the maximum great circle distance allowed from the Driver observation; p and z are the maximum pressure difference (in hPa) and height difference (in km), respectively, allowed between the Driver and Dependent observations. Note that the collocation tool opts to compute pressure differences over height differences: height differences are calculated only if the pressure variable is not available in either or both the Driver and Dependent datasets.

| Dataset | Δt | Δx | Δp | Δz |
|---|---|---|---|---|
| Aeolus | 60 minutes | 100 km | 0.04 log10(hPa) | 1 km |
| Aircraft | 60 minutes | 100 km | 0.04 log10(hPa) | 1 km |
| AMV | 60 minutes | 100 km | 0.04 log10(hPa) | 1 km |
| Loon | 60 minutes | 100 km | 0.04 log10(hPa) | 1 km |
| Sonde | 90 minutes | 150 km | 0.04 log10(hPa) | 1 km |

2.2.2. Plotting Tool

The application's plotting tool consists of visualization and statistical analysis functions developed for use in conjunction with the SAWC wind archive. The plotting tool uses the index files generated by the collocation tool to extract and compare matched wind data. The user can modify several parameters for the plotting tool (Table 6), including the option to super-ob the matches. In the context of SAWC, super-obbing refers to the averaging of all Dependent winds (from one Dependent dataset, e.g., all aircraft winds) that match the same Driver observation, resulting in one Dependent observation (representing the mean of the multiple collocations) per one Driver observation. Super-obbing is recommended in order to reduce processing time and produce smoother results.

**Table 6.** List of main input parameters the user can modify in the plotting tool.

| Parameter | Description |
|---|---|
| Start Date, End Date | Start date and end date (year, month, day, hour) over which to run the collocation tool. |
| Driver Dataset Name | Name of Driver dataset. |
| Dependent Dataset Names | Names of Dependent datasets to be compared to the Driver. |
| Path to Input Index Files | Full path to location where input collocation index files are located. |
| Path to Output Plots | Full path to location where output plots are to be saved. |
| Super-ob Choice | Choice to super-ob (average) multiple collocations per Driver observation, or use all collocations for statistical analysis. Options: -1 (use all collocations), or 0 (super-ob) |

Once the matched winds are extracted from the index files but before they are compared and assessed, the non-Aeolus winds are projected onto the Aeolus HLOS direction prior to any analysis only if Aeolus is included in the comparison (see Appendix A). Note that the HLOS winds have both a speed and direction, but the direction is constrained to be along the HLOS, and both are represented in SAWC by a signed speed with positive values away from the satellite.

After the HLOS conversion (if applicable), a second set of QCs is applied to all winds. This entails a general wind gross check where wind differences between the Driver and Dependent observations are rejected if they exceed |25| m/s. This threshold was chosen to omit extraneous outliers and follows other established collocation practices (e.g., Daniels et al., 2012). The gross check is applied to all dataset pairings selected for comparison, after which the plotting tool analyzes the QC'd matched winds and generates a host of figures that quantify the comparisons. The plotting tool computes global and regional statistics simultaneously, where regional refers to the Northern Hemisphere extratropics (NH), Tropics (TR), and Southern Hemisphere extratropics (SH), and the region borders are 30°N and 30°S. Seasonal statistics may be computed by rerunning the plotting tool for different date ranges. In addition, the plotting tool must be run separately for each Aeolus wind regime (RayClear or MieCloud).

The collocation and plotting tools are designed to be flexible and handle additional datasets not yet available in the archive. The user manual provides instructions on how to include such datasets for offline comparisons (Lukens et al., 2023b).

**3. Demonstrations of SAWC's Utility**

*3.1. Comparisons between Wind Datasets*

This section exemplifies the value of SAWC through a one-year global evaluation of Aeolus RayClear and MieCloud winds. Figure 3 shows mapped number densities or counts per grid cell of QC'd Aeolus winds for the study period (00 UTC September 1, 2019 through 18 UTC August 31, 2020) prior to collocation. Aeolus winds are observed on a global scale from the lower troposphere to the lower stratosphere. Higher MieCloud number densities highlight cloudy regions, while the number density for RayClear winds presents a more uniform distribution globally. The one-year evaluation is stratified by season (September-October-November (SON), December-January-February (DJF), March-April-May (MAM), and June-July-August (JJA)), geographic region (NH, TR, SH), and Aeolus wind regime (RayClear and MieCloud).

For the demonstration, the Driver dataset is Aeolus, and the Dependent datasets contain aircraft, sondes, Loon stratospheric balloons, and AMV wind observations. The winds are QC'd using the wind community-recommended parameters listed in Table 4, and the Dependent observations are collocated with the Driver observations using the default collocation criteria listed in Table 5. Prior to the analysis, the collocated Dependent winds are projected onto the Aeolus HLOS direction and are super-obbed for every Driver observation. The general wind gross check (see Section 2.2.2) is applied to all collocated pairs.

Mean statistics are computed for the collocation samples and include correlation (r) and root mean square differences (RMSDs) as defined in Wilks (2011), as well as mean differences in wind speed (Mean_Diffs) and standard deviations (SD) of those differences (SD_Diffs) as defined in Lukens et al. (2022) (notated in their paper as MCD and SDCD, respectively). The SD_Diffs and RMSDs tend to be very similar and as such any discussion involving SD_Diffs also applies to the RMSDs. In addition, the statistical significance of wind differences at the 95% confidence level (p-value < 0.05) is assessed using the paired two-tailed Student's *t*-test.

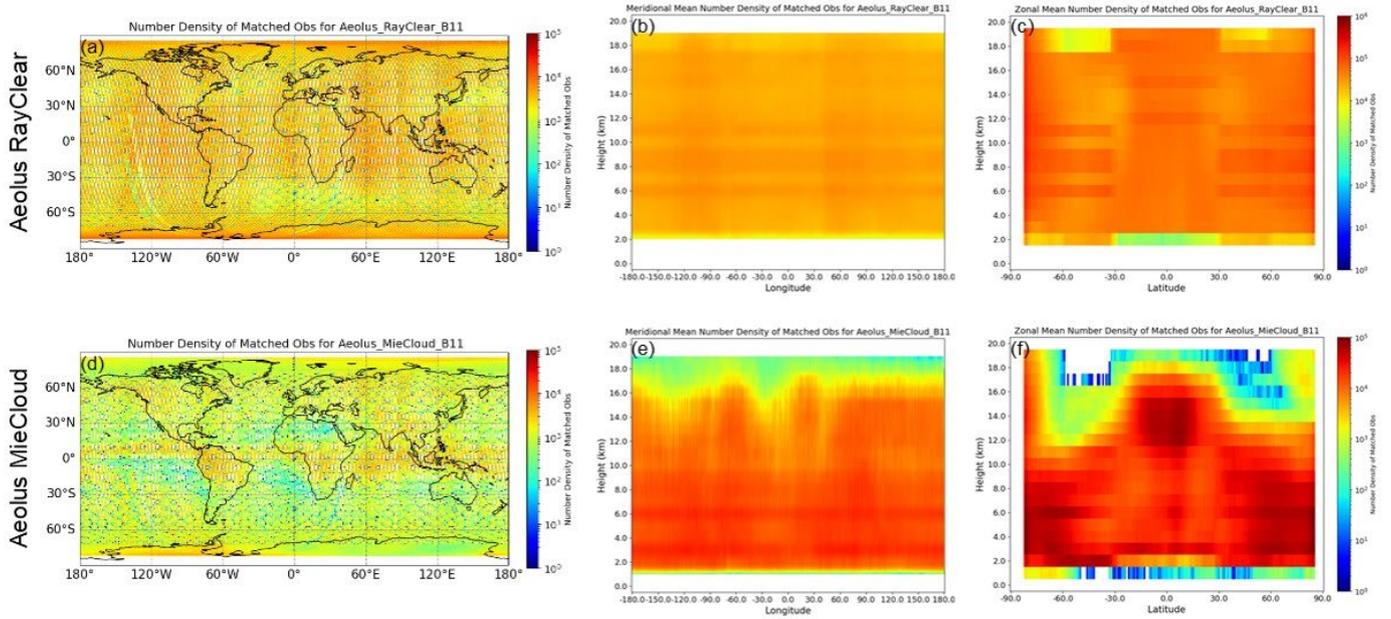

**Figure 3.** Observation number density maps on a latitude-longitude plane (left column), height-longitude plane (middle column), and height-latitude plane (right column) for QC'd Aeolus winds from the (**a-c**) Rayleigh-clear regime, and (**d-f**) Mie-cloudy regime for Sept 2019-Aug 2020, prior to collocation with the Dependent datasets. Colors indicate number density per grid cell, with dimensions of 1° x 1° for panels (**a,d**), and 1 km x 1° for the other panels.

3.1.1. Dataset Comparison Results

Figure 4 presents the number densities of collocated observation pairs between Dependent and Aeolus RayClear winds. Fig. 4a shows that aircraft winds matched with Aeolus cover much of the NH, with pockets of higher densities observed over the contiguous United States and Europe, as expected (e.g., James et al., 2020, their Fig. 5a). There are few collocations over the SH largely due to sparse spatial coverage of aircraft in the region, particularly over the oceans (see Fig. 2b). Many aircraft winds are matched with Aeolus at cruise altitude (8-12 km) in the upper troposphere (Figs. 4b-c), and also in the mid- and lower troposphere during the ascending and descending legs of each flight. AMVs exhibit near-global coverage (Fig. 4d) and are collocated with Aeolus throughout the vertical between the lower troposphere (~800 hPa) and the lower stratosphere (~100 hPa) (Figs. 4e-f). There is a noticeable dip in numbers near 65 degrees of latitude between the regions of excellent coverage for GEO and LEO observing systems. Sonde winds are generally collocated with Aeolus over land, particularly in the NH, between the lower troposphere (800 hPa) and the stratosphere (above 100 hPa) (Figs. 4g-i). Note that the horizontal bars of varying density displayed in Figs. 4e-f and 4h-i are a result of collocating Dependent winds with Aeolus at Aeolus pressure levels rather than heights. Aeolus winds are each assigned both a height and pressure, but because they are vertically sampled based on height, and height is linear in the vertical while pressure is nonlinear, Aeolus winds have a more uniform distribution of number density on a height-latitude/-longitude plane (see Fig. 3) relative to a pressure-latitude/-longitude plane. Loon winds are collocated with Aeolus in the lower stratosphere in the TR and part of the SH (Figs. 4j-l) during the collocation period; note that the 3D distribution of Loon matches may vary depending on the time period chosen. Aeolus MieCloud collocations exhibit similar spatial distributions (not shown), except that aircraft and sondes total somewhat fewer collocations, and AMVs have more collocations than their RayClear counterparts. Additionally, there are no collocations between MieCloud and Loon as Loon observes winds at levels above the MieCloud observations.

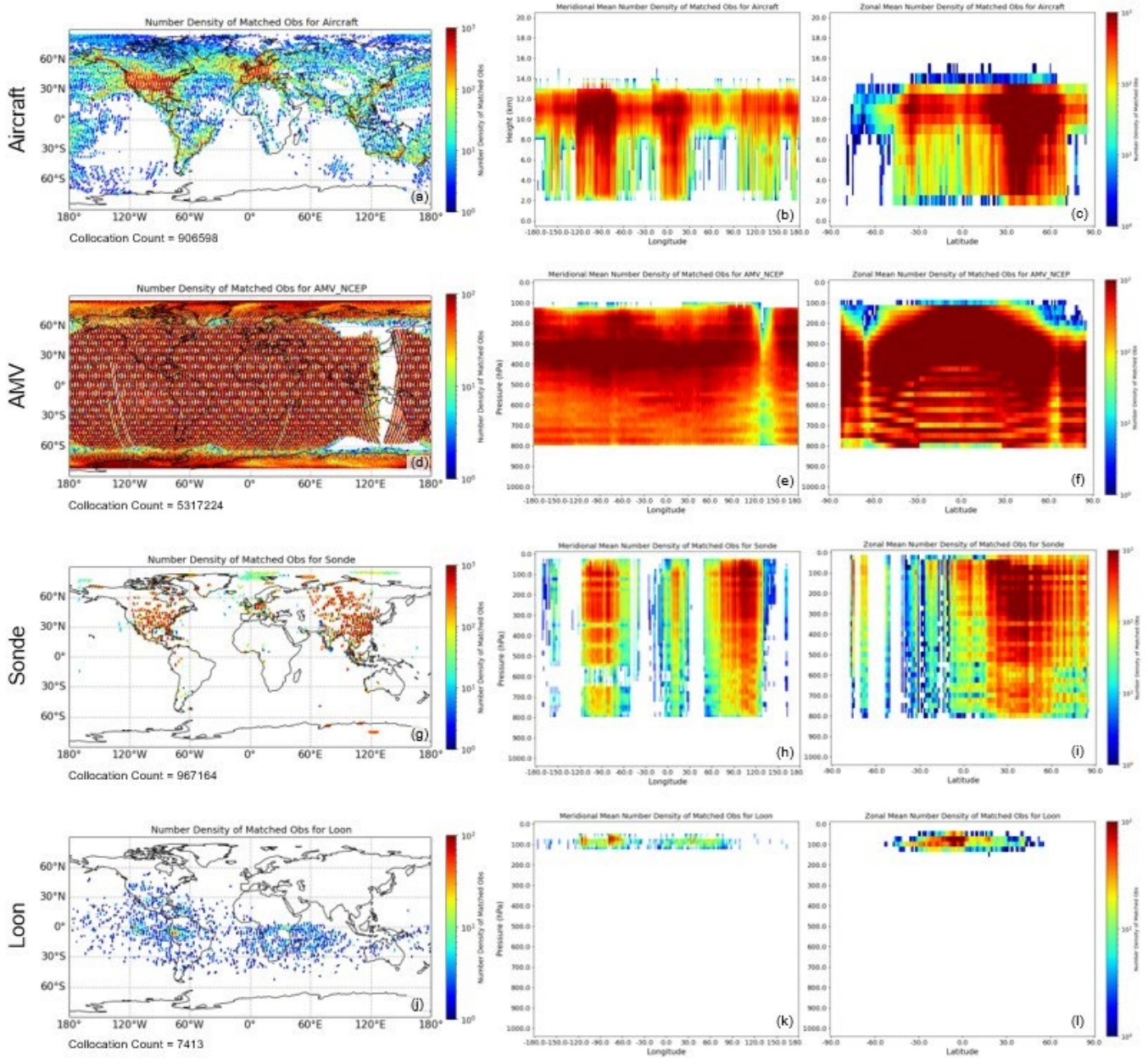

**Figure 4.** Observation number density maps on a latitude-longitude plane (left column), height/pressure-longitude plane (middle column), and height/pressure-latitude plane (right column) for (**a-c**) aircraft winds, (**d-f**) AMVs, (**g-i**) sonde winds, and (**j-l**) Loon winds, all collocated with Aeolus Rayleigh-clear winds for Sept 2019-Aug 2020. Colors indicate number density per grid cell, with dimensions of 1° x 1° for panels (**a,d,g,j**), 1 km x 1° for panels (**b,c**), and 25 hPa x 1° for the other panels.

Figure 5 presents density scatterplot comparisons of global HLOS winds matched between the Dependent datasets (y-axis) and the Driver (x-axis) (Aeolus RayClear in the top row, MieCloud in the bottom row) for the collocation period. In general, all four Dependent datasets match well with both Aeolus wind regimes, as illustrated by the observations falling near the one-to-one line that indicates a perfect match. MieCloud comparisons have higher correlations and lower SD_Diffs than RayClear, consistent with the general higher accuracy of MieCloud winds, in agreement with the findings from Abdalla et al. (2021). AMVs, aircraft, and sonde winds are all highly correlated (>0.9) with Aeolus (Figs. 5a-f), with aircraft exhibiting statistically significant differences from both Aeolus wind regimes, and AMVs being significantly different from RayClear winds. Loon winds exhibit the lowest correlation (0.84) and largest Mean_Diff (|0.88| m/s) and SD_Diff (7.4 m/s) of the RayClear comparisons. This is likely influenced by Loon's comparatively small

sample size (<8000 for Loon and >900,000 for the others) (Fig. 5g) coupled with fewer Aeolus observations in the stratosphere, thus resulting in few chances to match Loon with Aeolus.

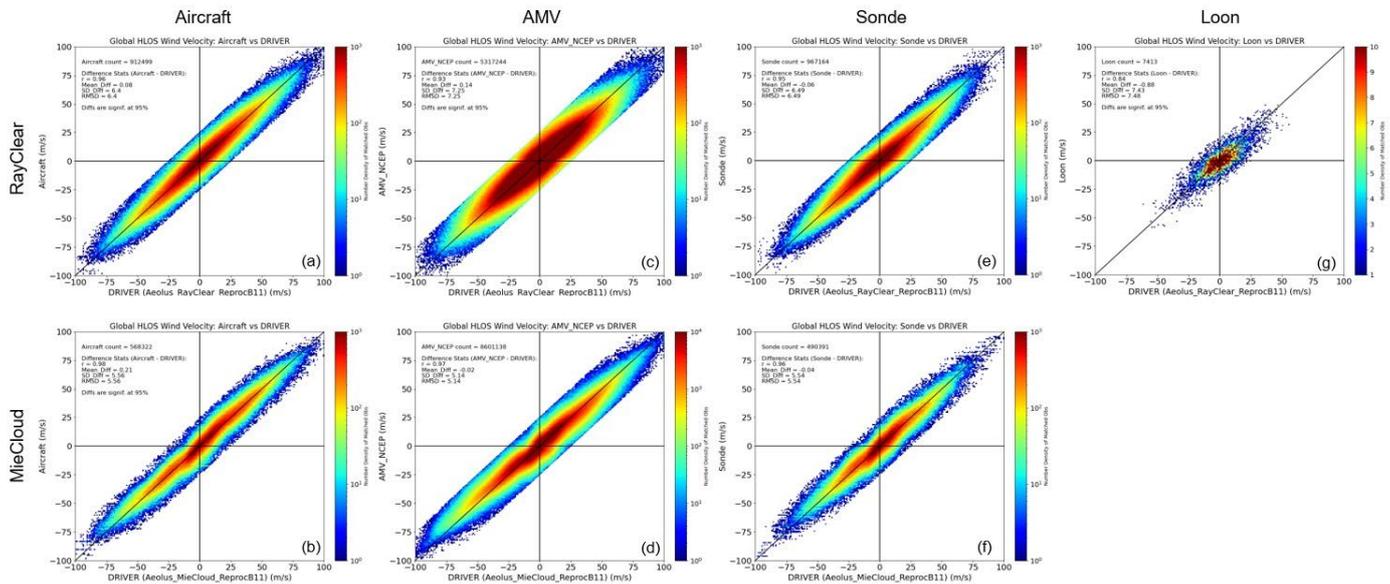

**Figure 5.** Density scatterplots of HLOS wind differences for RayClear (top row) and MieCloud comparisons (bottom row) with aircraft, AMVs, sondes, and Loon for Sept 2019-Aug 2020. Mean statistics are displayed in the top left corner of each panel and include collocation counts, correlation (r), mean HLOS wind difference (Mean_Diff) in m/s, standard deviation of Mean_Diffs (SD_Diff) in m/s, and RMSD in m/s. Colors indicate number density per 1 m/s x 1 m/s cell. If the wind differences are statistically significant at the 95% level, this is stated below the statistics.

Time series of global wind comparisons are illustrated in Fig. 6. With the exception of Loon, RayClear Mean_Diffs are generally small throughout the study period (Fig. 6a), and correlations are above 0.9 until June 2020 where they decrease slightly toward 0.8. This decrease corresponds to an increase in Mean_Diff magnitudes and SD_Diffs with time, as well as a decrease in the number of observations, all of which are also observed when comparing RayClear winds to the ECMWF model (Abdalla et al., 2021). The increase in SD_Diffs can be explained by the observed signal loss along the instrument's atmospheric path over time (Abdalla et al., 2021). Loon comparisons show the largest Mean_Diffs and SD_Diffs throughout the collocation period, as well as the lowest correlations and collocation counts. MieCloud comparisons exhibit small Mean_Diffs, higher correlations, and smaller SD_Diffs (Fig. 6b) compared to RayClear. Note that the MieCloud Mean_Diffs increase with time (Fig. 6b, top panel). This increase is likely attributed to the uneven distribution of cloud cover between Aeolus ascending vs. descending orbits (i.e., more clouds due to increased convection during ascending orbits) (Abdalla et al., 2021).

Mean vertical profiles of Mean_Diffs and SD_Diffs for the study period are stratified per geographic region in Fig. 7. RayClear Mean_Diffs are generally < |1.0| m/s and are statistically significant throughout the vertical in each region (Figs. 7a-c). The largest Mean_Diffs are observed in the mid- to lower troposphere (below 500 hPa) where the collocation counts are small. SD_Diffs are steady at 6-7 m/s throughout most of the atmospheric column and increase slightly above 200 hPa that could be attributed to larger observation errors in the presence of high winds and stronger vertical wind shear, e.g., for AMVs (Cotton et al., 2020). MieCloud comparisons (Figs. 7d-f) show smaller Mean_Diffs (< |0.5| m/s) and SD_Diffs (5-6 m/s) in the NH and TR. The notable exception is in the SH where the Mean_Diffs are larger for all Dependent datasets (≥ |1.0| m/s) and become larger with height. The corresponding SD_Diffs increase with height and are likely attributed to larger observation errors at upper levels due in part to a smaller sample size combined with high vertical wind shear (e.g., Bormann et al., 2002; Cordoba et al., 2017).

Figure 8 displays the dependence of Mean_Diffs and SD_Diffs on Aeolus wind speed during the study period. In general, Mean_Diffs tend to become larger for faster Aeolus winds for both Aeolus wind regimes. RayClear comparisons (Figs. 8a-c) have larger SD_Diffs that tend to increase with faster Aeolus winds relative to MieCloud (Figs. 8d-f), again highlighting the general higher accuracy of MieCloud winds. This is particularly evident for AMVs (Figs. 8a-c, green dashed lines) and in the SH where there are fewer observations. Further, the SD_Diffs have similar values to the SDs of the Dependent datasets, implying that the Dependent wind quality governs the SD_Diffs. In addition, the collocation counts for all datasets peak to the right where Aeolus HLOS velocities are positive, implying that many Dependent winds match Aeolus at upper levels where westerly flow dominates. This is supported by the collocation counts displayed in Fig. 7. The shift in peak counts exemplifies SAWC's value in revealing the unique behavior of the data.

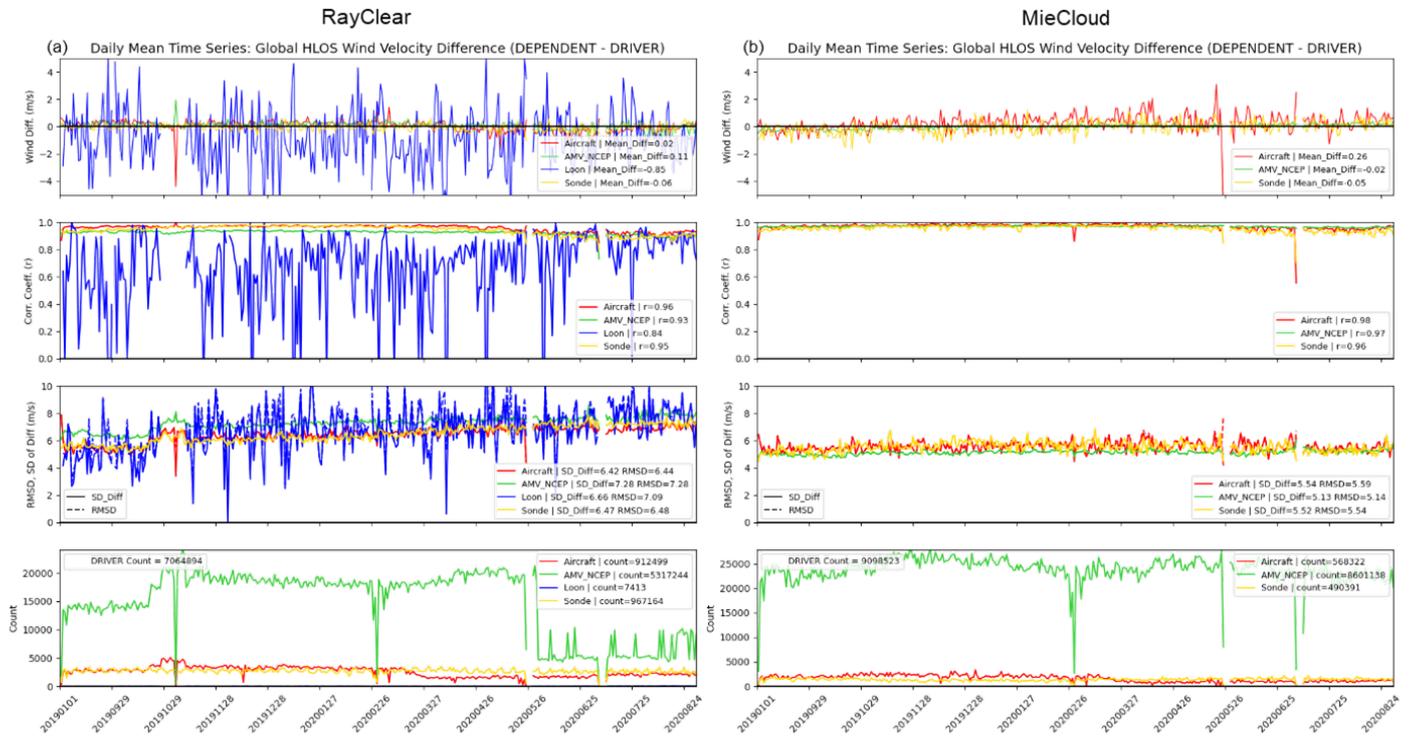

**Figure 6.** Time series of Mean_Diffs in m/s (row 1, i.e., the top row), correlation coefficients (r) (row 2), RMSD and SD_Diffs in m/s (row 3), and collocation counts (row 4, i.e., the bottom row) for (**a**) RayClear, and (**b**) MieCloud comparisons during Sept 2019-Aug 2020. Mean statistics are displayed in the legends for each panel. Colors denote each Dependent wind dataset.

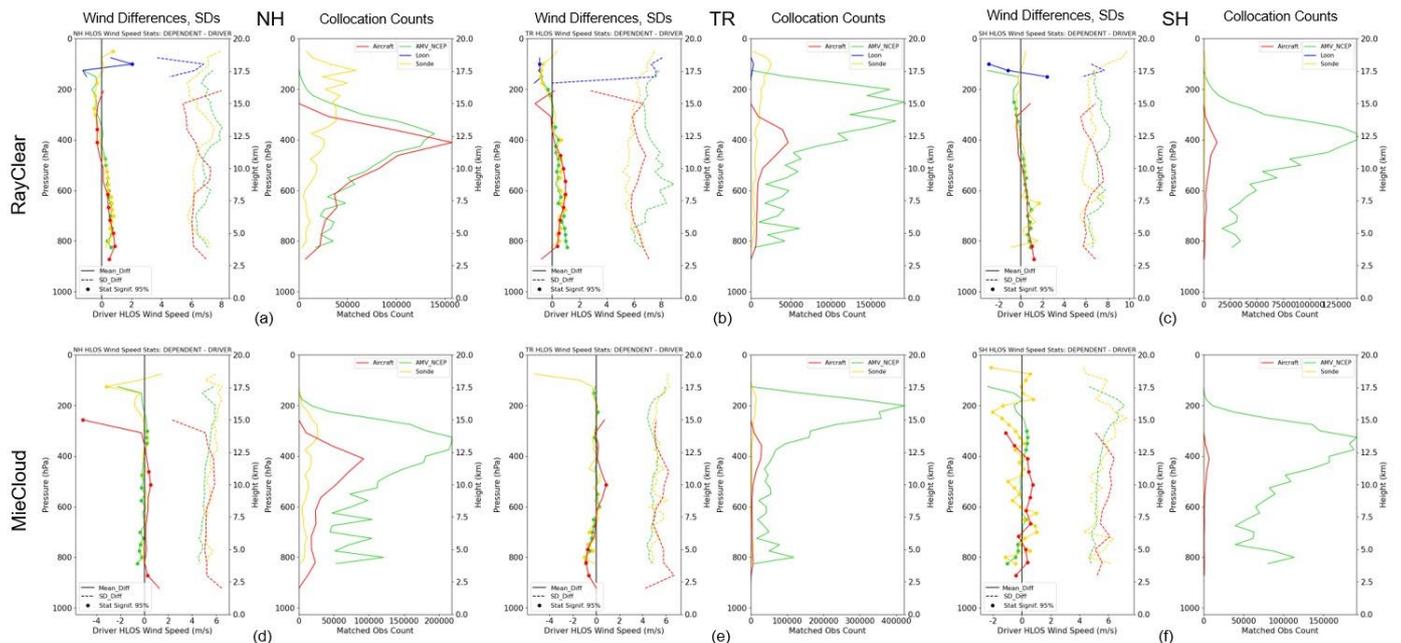

**Figure 7.** Vertical profiles of Mean_Diffs (solid lines, left panels) and SD_Diffs (dotted lines, left panels) per height/pressure level comparing the Dependent datasets (colors) with Aeolus RayClear (top row) and MieCloud winds (bottom row) for the NH (left group), TR (center group), and SH (right group) during Sept 2019-Aug 2020. Corresponding collocations counts are displayed in the right panels of each group. Solid dots indicate statistically significant Mean_Diffs at the 95% level.

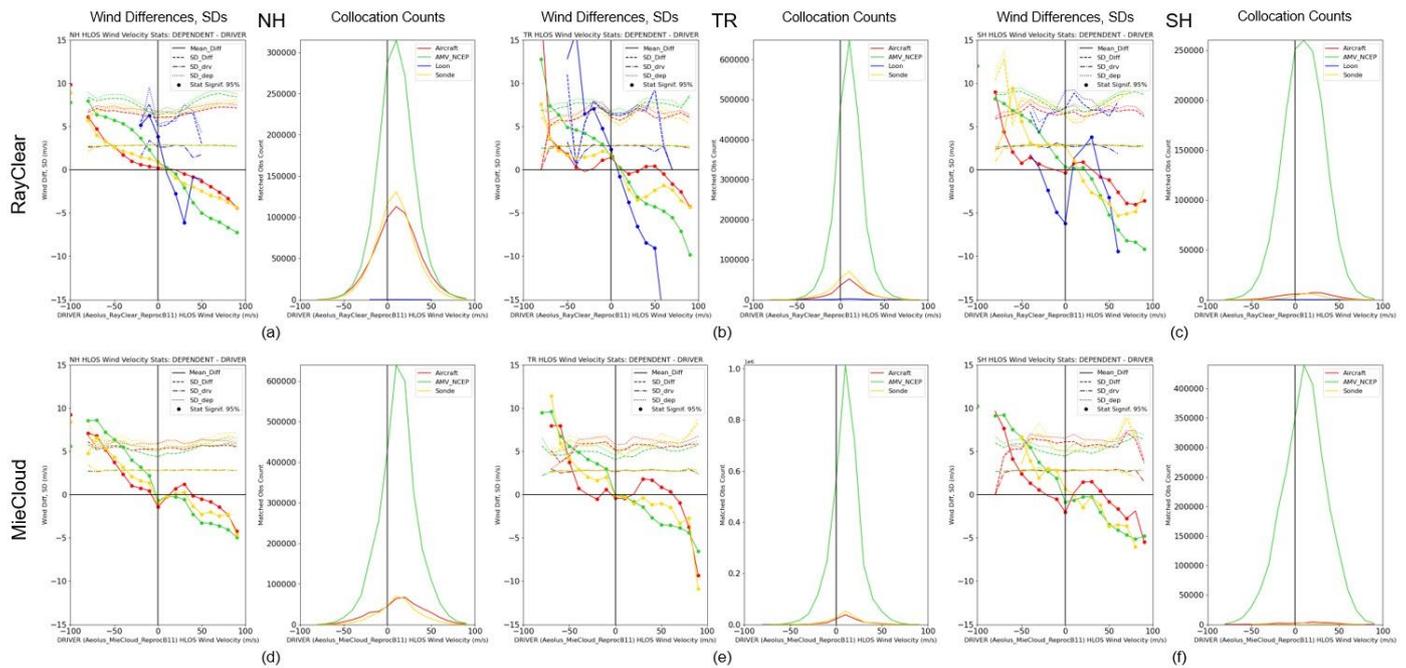

**Figure 8.** Mean_Diffs (solid lines, left panels), SD_Diffs (dotted lines, left panels), as well as SDs of the Driver (dash-dot lines, left panels) and SDs of the Dependent datasets (dotted lines, left panels) per Driver wind speed binned by 10 m/s comparing the Dependent datasets (colors) with Aeolus RayClear (top row) and MieCloud winds (bottom row) for the NH (left group), TR (center group), and SH (right group) during Sept 2019-Aug 2020. Corresponding collocations counts are displayed in the right panels of each group. Solid dots indicate statistically significant Mean_Diffs at the 95% level.

### 3.1.2. AMV Wind Types vs. Aeolus

A unique capability of the SAWC plotting tool is the intercomparison of four main AMV wind types: infrared (IR), visible, water vapor cloudy sky (WVcloud), and water vapor clear sky (WVclear). AMV type comparisons are automatically performed only if the user selects an AMV dataset for analysis. Global and regional statistics are computed simultaneously. Figure 9 presents global density scatterplots of each AMV type relative to Aeolus RayClear (top row) and MieCloud winds (bottom row) for the collocation period. All AMV types match well with Aeolus, with each comparison exhibiting correlations > 0.9, and MieCloud comparisons exhibiting lower SD_Diffs (5-6 m/s) and higher correlations (> 0.94) than RayClear, in agreement with the findings of Lukens et al. (2022). The Mean_Diffs are near zero except for comparisons of winds from different scenes, e.g., IR vs. RayClear (Mean_Diff is 0.39 m/s) (Fig. 9a) characterizes the collocation between cloud-tracked AMVs and Aeolus clear scene winds, and WVclear vs. MieCloud (Mean_Diff is -0.66 m/s) (Fig. 9h) represents the collocation between clear-sky AMVs and Aeolus cloudy scene winds.

Mean vertical profiles of Mean_Diffs (AMV type minus Aeolus) and the corresponding SD_Diffs are shown per geographic region for the study period in Fig. 10. RayClear comparisons exhibit similar Mean_Diff profiles in each region: larger, statistically significant differences (1-2 m/s) in the mid- to lower troposphere where there are fewer collocations, and smaller statistically insignificant differences (< |1| m/s) at pressure levels < 400 hPa where the collocation counts are higher. SD_Diff profiles of IR and visible AMVs exhibit similar behavior: they are steady at ~6 m/s in the lower troposphere, increase with height until 400 hPa where they peak at 8 m/s, and decrease at upper levels. Likewise, WVcloud and WVclear SD_Diff profiles are similar in that they range from 8 to 10 m/s and tend to decrease with height throughout the troposphere.

Unlike RayClear, MieCloud profiles exhibit different behavior in each geographic region. Mean_Diffs in the NH (Fig. 10d) and TR (Fig. 10e) are larger (-1 to -2 m/s) and statistically significant in the mid- to lower troposphere where there are fewer collocations. In the upper troposphere (pressures < 400 hPa), the Mean_Diffs are near zero and not statistically significant except for WVcloud in the NH. MieCloud SD_Diff profiles exhibit similar behavior to their RayClear counterparts: IR and visible SD_Diffs are relatively constant throughout most of the atmospheric column, while WVcloud and WVclear profiles exhibit larger values at lower levels and decrease at a faster rate with height. This is particularly evident in the TR (Fig. 10e). In the SH (Fig. 10f), MieCloud Mean_Diffs are relatively small (around |1| m/s) and statistically significant throughout the vertical. SD_Diffs range from 5 to 7 m/s at lower levels and increase with height for all AMV types, particularly at pressures < 400 hPa. This behavior is consistent with the results discussed in Lukens et al. (2022) and could be attributed to height assignment errors near jet levels, as AMV uncertainties tend to increase in areas with higher wind speed and stronger shear (Posselt et al., 2019; Bormann et al., 2002; Cordoba et al., 2017).

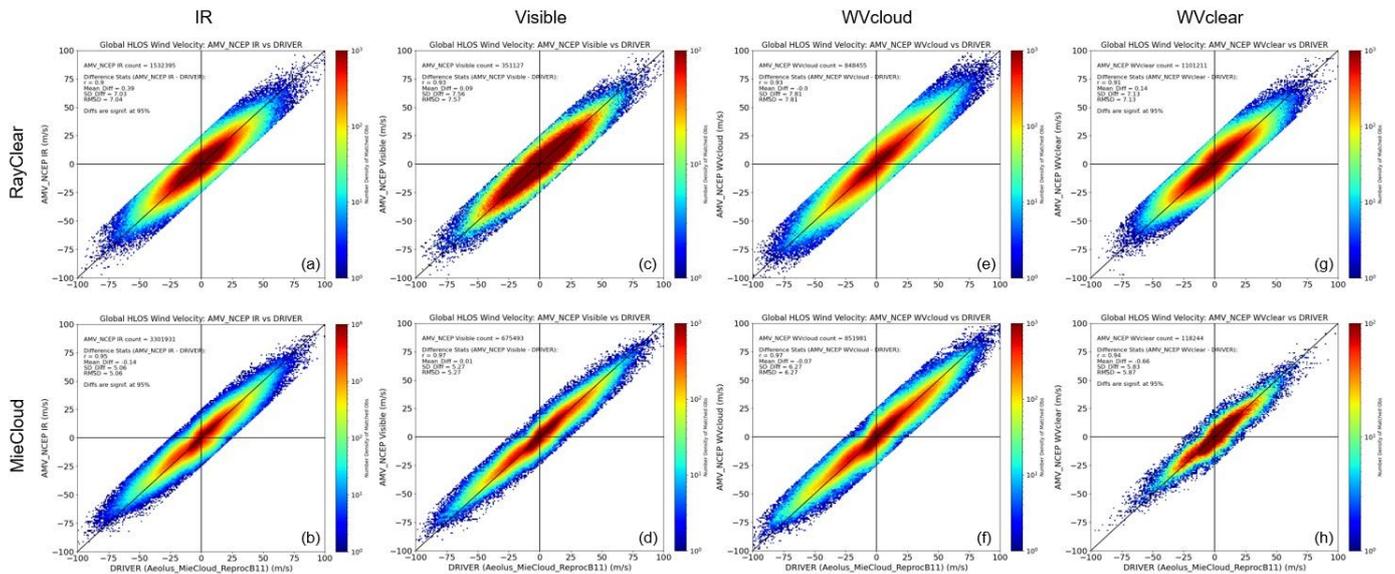

**Figure 9.** As in Fig. 4 but comparing Aeolus with IR, visible, WVcloud, and WVclear AMVs.

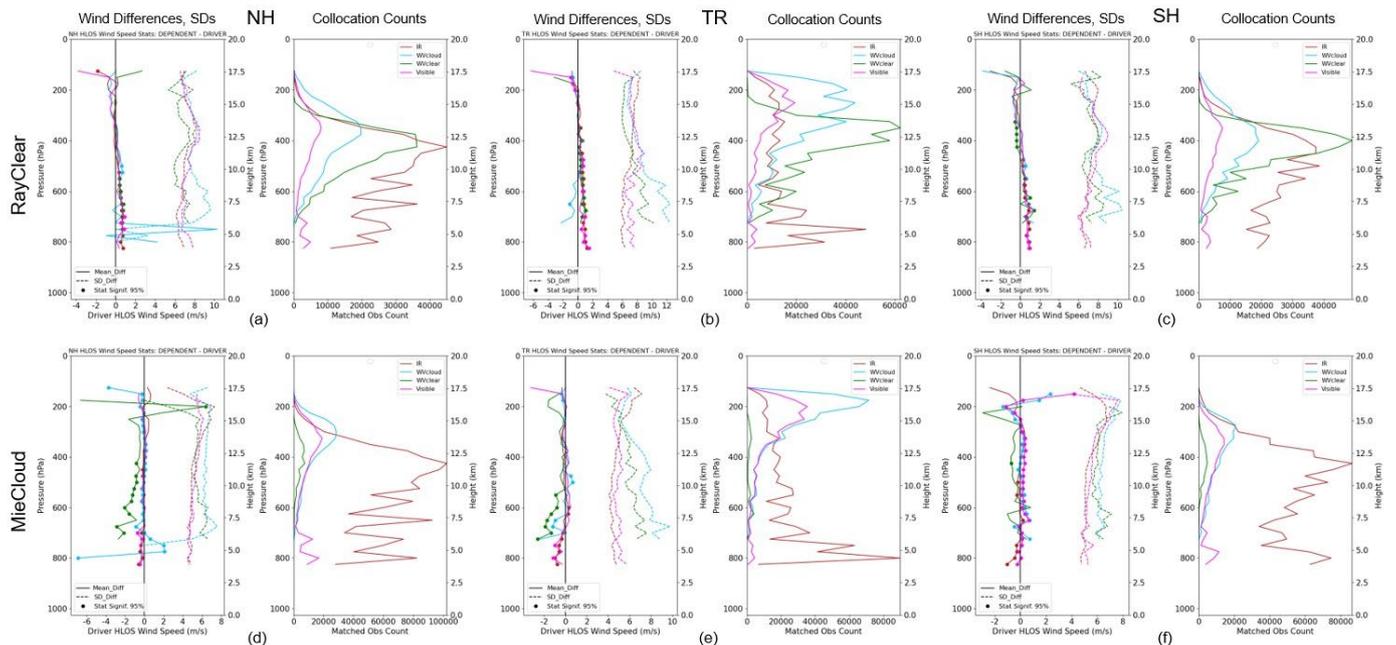

**Figure 10.** As in Fig. 6 but comparing Aeolus RayClear winds with IR, visible, WVcloud, and WVclear AMVs.

3.1.3. Aircraft vs. Aeolus during the COVID-19 Pandemic

SAWC is capable of providing insight into the environmental impact of unprecedented global events, such as the COVID-19 pandemic. The COVID-19 pandemic effectively shut down the globe beginning in March 2020. Several facets of the global economy were heavily impacted, particularly air travel which was considerably reduced. Studies have since examined the pandemic's effect on aircraft observations during this time and the subsequent impact of the reduction in aircraft data on the climate, atmospheric composition, and NWP (James et al., 2020; Quaas et al., 2021; Reifenberg et al., 2022; Clark et al., 2021; Chen, 2020; Rani et al., 2023).

Figure 11 displays 3D map projections of aircraft winds collocated with Aeolus RayClear winds for each 3-month season during the collocation period. Prior to the onset of the pandemic (SON and DJF; Figs 11a and 11b, respectively), the number of collocations per season remained relatively constant around 290,000 with many observed in the NH and upper troposphere, as expected. Wind profiles are clearly visible across the global domain, and pockets of high wind speed are observed in the extratropics at jet level, again as expected. After the pandemic began (MAM and JJA; Figs 11c and 11d, respectively), the total number of collocations decreased by over 50%, reflecting the substantial aircraft data gap that emerged as a result of the pandemic's impact on air travel (James et al., 2020; Chen 2020). This is particularly evident during JJA in the TR and SH where the number of aircraft wind profiles is largely reduced. Similar results are found for MieCloud comparisons (not shown).

Figure 12 presents an examination of the pandemic's impact on aircraft data quality. SD_Diffs of aircraft versus Aeolus RayClear winds on a height-latitude plane are shown in Figs. 12a-d, and corresponding number densities or counts per grid cell are displayed in Figs. 12e-h. Before the pandemic (SON and DJF), the SD_Diffs are generally 5-8 m/s in the mid- to upper troposphere (6-12 km) and throughout the NH where the majority of winds are collocated. During the pandemic (MAM and JJA), the SD_Diffs increased at upper levels (8-12 km) where the number of collocations decreased, particularly in the NH in JJA. MieCloud comparisons exhibit similar patterns (not shown). The results support that the reduction in the number of commercial aircraft flights during the pandemic contributed to poorer quality observations, which in turn have been found to have a considerable negative impact on regional NWP forecast skill (James et al., 2020; Chen, 2020; Moninger et al., 2010; Petersen, 2016).

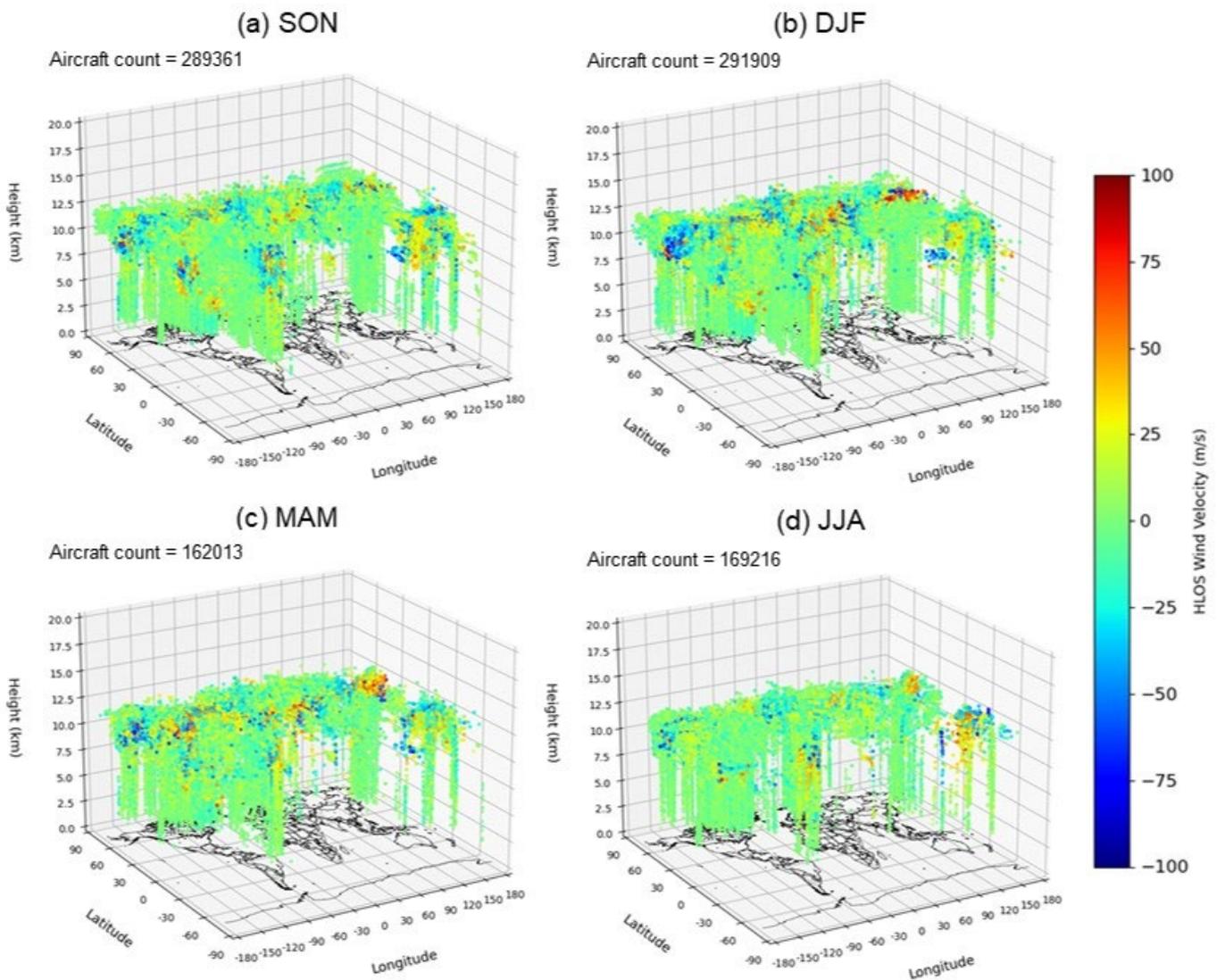

**Figure 11.** 3D map projections of aircraft winds collocated with Aeolus RayClear winds during Sept 2019-Aug 2020, stratified by 3-month seasons (**a**) SON, (**b**) DJF, (**c**) MAM, and (**d**) JJA. Colors indicate the aircraft HLOS wind velocity in m/s. Corresponding collocation counts are displayed in the top left corner of each panel.

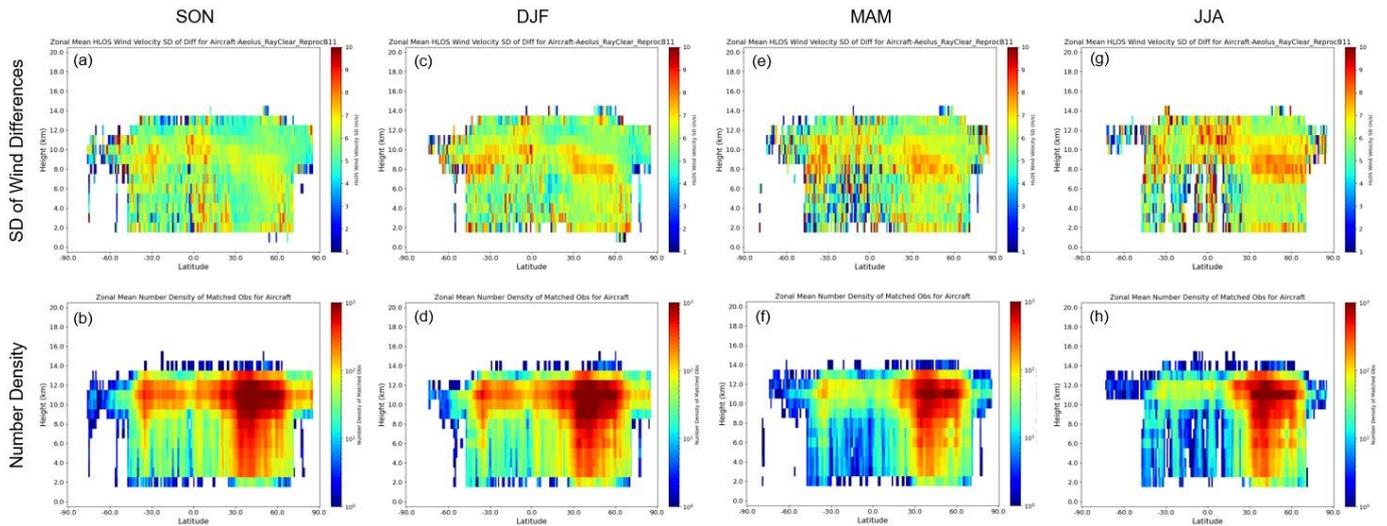

**Figure 12.** Height-latitude plots of aircraft vs. Aeolus RayClear SD_Diffs in m/s (top row) and corresponding collocation counts (bottom row) during Sept 2019-Aug 2020, stratified by 3-month seasons (**a**) SON, (**b**) DJF, (**c**) MAM, and (**d**) JJA. Colors indicate (top row) aircraft HLOS wind velocity in m/s, and (bottom row) observation number density. Each grid cell is 1 km x 1°.

*3.2. Observation Error Estimation for DA*

SAWC's utility extends beyond wind comparison statistics to, e.g., observation error estimation that is critical for DA. For example, SAWC can be used to investigate how well DA observation errors match actual observation collocation difference standard deviations. An example of this is presented in Fig. 13 that shows global mean observation error estimates per Aeolus MieCloud wind speed bin during the collocation period, based on the Desroziers method (Desroziers et al., 2005). The variance for Aeolus (dash-dot lines) is computed using observation error values provided by ESA. The variance for each Dependent dataset (dotted lines) is computed using input observation errors in NOAA operations. The total variance (solid lines) equals the Aeolus variance plus the Dependent variance. The square of the mean wind difference (Dependent minus Driver) that is computed in SAWC is displayed as dashed lines. Ideally, the total variance and the square of the difference should be nearly identical, implying that the observation error covariance estimates in DA match the observations. For aircraft and sondes, this is the case for slower winds (< |25| m/s) when using a single observation error value (3.0 m/s) for all winds in each dataset.

AMV error covariance estimation is challenging. It is clear in Fig. 13a that the use of a single observation error value (10 m/s) is not ideal, as the total variance and square of the difference differ by ~10-100 m/s. When using a range of vertically-varying errors (Fig. 13b), the results improve, as the shape of the total AMV variance is similar to that of the square of the mean wind difference; however, their magnitudes still greatly differ for most observations. Further investigation into AMV error covariance characterization is out of the scope of this work; however, it is necessary and strongly recommended for future studies.

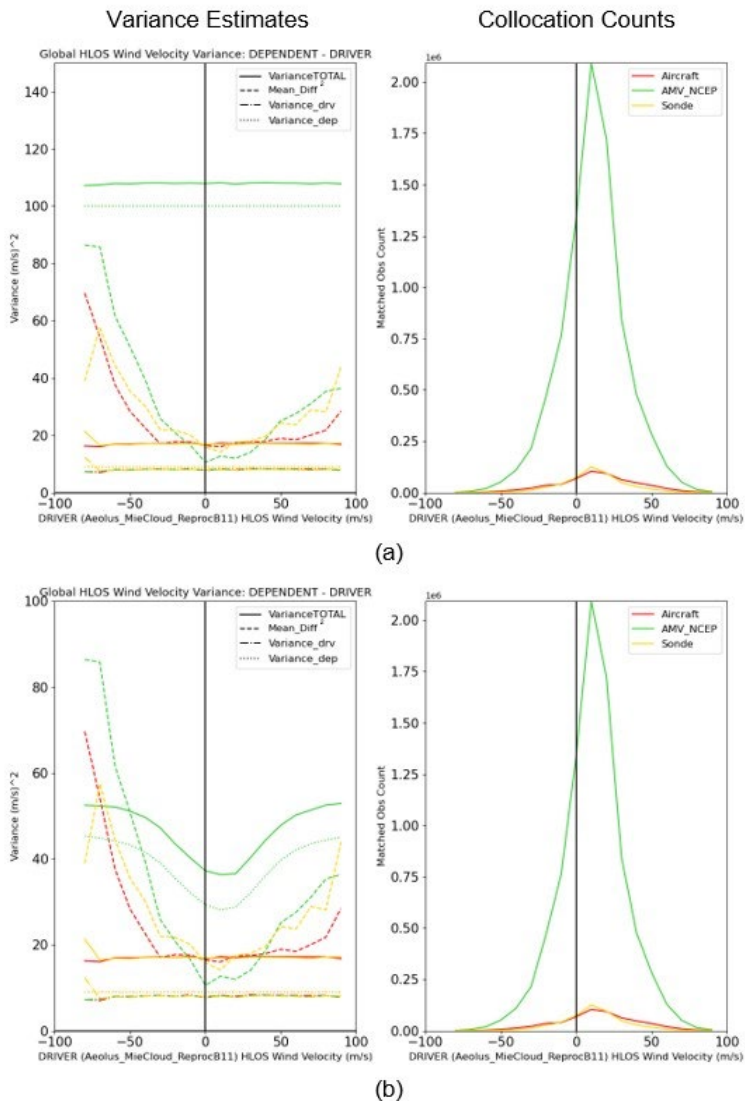

**Figure 13.** (left column) Global mean observation error variance estimates vs. Driver wind speed binned by 10 m/s comparing the Dependent datasets (colors) with Aeolus MieCloud winds during Sept 2019-Aug 2020, and (right column) corresponding collocation counts. For variance estimates: Error variance of the Driver (dash-dot lines) using observation error values from the data producers, error variance of the Dependent datasets (dotted lines) using observation errors that are used as input in NOAA operations, total variance that equals the Driver variance plus the Dependent variance (solid lines), and the square of the mean wind difference (Dependent-Driver) computed in SAWC (dashed lines). In (**a**), Dependent observation errors in each dataset are set to a single value for all winds: Aircraft = 3.0 m/s, Sonde = 3.0 m/s, AMV = 10 m/s. In (**b**), aircraft and sonde errors are the same as in (**a**), but the AMV observation error is set to a range of vertically-varying values from 3.8 m/s at pressures > 1000 hPa to 7.0 m/s at pressures < 250 hPa, and represents the median range of input observation errors used for each satellite in NOAA operations.

## 4. Summary and Discussion

This article introduces the System for Analysis of Wind Collocations (SAWC) that was jointly developed by NOAA/NESDIS/STAR, UMD/CISESS, and UW-Madison/CIMSS to support users' needs to acquire and analyze wind data from multiple sources, and to further address the requirement for highly accurate 3D winds put forth in the National Academies' 2017-2027 decadal survey. SAWC's end-to-end process can be described in three steps (Fig. 1):

1. Data Acquisition, where wind observations are acquired from aircraft, satellites (Aeolus winds and AMVs), sondes, and stratospheric superpressure balloons (Fig. 2), converted to a common format (netCDF), and archived;
2. Collocation of Winds, where users can utilize the SAWC collocation tool developed for their intercomparison to produce a collection of matched winds between different datasets; and
3. Analysis and Visualization, where users can interact with the SAWC plotting tool to visually and statistically compare the matched winds based on their research needs.

The collocation and plotting tools are flexible and designed to handle additional wind datasets not yet available in the archive, e.g., radial winds. Since radial winds are similar in form to Aeolus winds in that the winds are retrieved along one direction, the Aeolus HLOS projection algorithm could also be applied to radial winds (see Appendix A). At present, the SAWC collocation and plotting tools can only handle point observation wind datasets. Gridded datasets (e.g., NWP reanalyses) as well as other observational products such as cloud optical and physical parameters are being considered for inclusion in future iterations of SAWC. SAWC is publicly available and is hosted online by NOAA/NESDIS/STAR at https://www.star.nesdis.noaa.gov/sawc.

A one-year evaluation of Aeolus winds was conducted to demonstrate the utility of SAWC. Aeolus (i.e., the Driver dataset) was compared to four other wind datasets available in the SAWC archive at the time of writing: AMV, aircraft, sonde, and Loon stratospheric balloon winds (Dependent datasets). The matched winds have regional to near-global spatial coverage that varies for each Dependent dataset (Figs. 3-4). Statistical comparisons highlight that the Dependent datasets match well with both Aeolus Rayleigh-clear and Mie-cloudy wind regimes for the study period (Fig. 5), with Mie-cloudy comparisons exhibiting smaller standard deviations of wind differences (SD_Diffs) relative to Rayleigh-clear, which is consistent with the higher accuracy of Mie-cloudy winds (Abdalla et al., 2021). SD_Diffs for Rayleigh-clear comparisons increase over time due to the degradation in the Aeolus retrieval signals (Abdalla et al., 2021) (Fig. 6). Mean wind differences (Mean_Diffs) in the NH and TR are relatively small throughout the vertical for both Aeolus wind regimes (< |1.0| m/s), and SD_Diffs remain relatively constant (5-7 m/s) (Fig. 7). SH Mean_Diffs are larger and corresponding SD_Diffs generally increase with height, as well as for faster winds (Fig. 8).

In addition, SAWC was used to compare individual AMV wind types to Aeolus (Figs. 9-10). MieCloud comparisons with infrared (IR), visible, water vapor cloudy sky (WVcloud), and water vapor clear sky (WVclear) AMVs each show lower SD_Diffs (5-6 m/s) and higher correlations (> 0.94) than RayClear, in agreement with the findings in Lukens et al., 2022. In the vertical, SD_Diffs for IR and visible AMVs remain relatively constant, while the WVcloud and WVclear profiles show larger values that tend to decrease with height, except in the SH where all SD_Diff profiles generally increase with height.

The collocation period chosen afforded the unique opportunity to showcase SAWC's capability in providing insight into aircraft wind observation data quality after the onset of the COVID-19 pandemic. A seasonal analysis highlights the reduced 3D spatial coverage of aircraft observations after the pandemic began in March 2020 (Fig. 11). SD_Diffs between aircraft and Aeolus winds increased at upper levels where the observation counts were noticeably reduced (Fig. 12), implying a degradation in data quality. The results support that the dearth of available aircraft observations during the pandemic contributed to poorer data quality, in agreement with the findings in James et al. (2020), and Chen (2020). In addition, to demonstrate SAWC's utility beyond intercomparisons, a brief example of how SAWC could be used to support observation error estimation in DA systems was presented (Fig. 13).

The potential value of SAWC is wide-ranging, from product validation and observation error characterization to advancements in the global Earth observing architecture. Moreover, SAWC could also be employed to test and establish new wind collocation standards, as has been recommended in order to keep up with advances in wind retrieval technologies and numerical weather prediction systems (e.g., AMV algorithms improvements, higher resolution observations, and advanced global DA systems).

**Author Contributions:** KG and KI proposed the project as co-investigators and provided the expertise that guided this work. BH and DS developed the collocation algorithm used. BH built the foundation for the SAWC collocation software application. KEL performed most of the work that included comparison analyses and major updates to the application's collocation and plotting tools. DH contributed a substantial upgrade to the collocation tool that greatly improved its cost efficiency. DS, RNH, and HL provided additional intellectual support that considerably improved the article. KEL prepared the paper, with contributions from all coauthors.

**Funding:** This research was supported by the NOAA/NESDIS Office of Projects, Planning, and Acquisition (OPPA) Technology Maturation Program (TMP) through the Cooperative Institute for Climate and Satellites (CICS) and the Cooperative Institute for Satellite Earth System Studies (CISESS) at the University of Maryland (UMD)/Earth System Science Interdisciplinary Center (ESSIC) (grant nos. NA14NES4320003 and NA19NES4320002) and the Cooperative Institute for Meteorological Satellite Studies (CIMSS) at UW-Madison (grant no. NA20NES4320003).

**Data Availability Statement:** All data used in this work are publicly available at https://www.star.nesdis.noaa.gov/data/sawc.

**Acknowledgments:** The authors thank ESA, ECMWF, the NCEP/EMC Obsproc team, and Loon, LLC for providing the data; Peter Marinescu for his contributions towards converting Aeolus BUFR data into 6-hour NCEP prepBUFR for the purpose of data assimilation in NOAA NWP; Chia Moeller for her contribution towards the initial development of tools for analysis and visualization; Iliana Genkova, Jim Jung, Jaime Daniels, Cathy Thomas, and Emily Liu for their expertise on NCEP prepBUFR and winds in NOAA NWP; the Aeolus Cal/Val teams, Mike Hardesty, Erin Jones, Chris Barnet, and Sebastian Bley for their scientific and technical expertise; and Sid Boukabara for his leadership and support of this project. The authors thank the administrators of the S4 supercomputer (Boukabara et al., 2016) at UW-Madison's Space Science Engineering Center (SSEC) who provided the groundwork for SAWC's initial development. The authors also thank the IT administrators at the NOAA/NESDIS/Center for Satellite Applications and Research (STAR) for their continued support in hosting and maintaining SAWC at STAR.

**Conflicts of Interest:** The authors declare no conflict of interest. The funders had no role in the design of the study; in the collection, analyses, or interpretation of data; in the writing of the manuscript; or in the decision to publish the results.

## Appendix A

If Aeolus is included in the wind comparison as either the Driver or a Dependent dataset, the non-Aeolus winds that are matched with Aeolus are projected onto the Aeolus HLOS wind direction prior to any statistical analysis or plotting taking place. If Aeolus is the Driver, all Dependent winds are projected onto the HLOS direction, and all statistical analyses are computed using the HLOS-projected winds. If Aeolus is listed as a Dependent dataset, the matched Driver winds are projected onto the HLOS direction, and statistical analyses only concerning Aeolus are computed using the HLOS-projected winds; otherwise, the Driver-Dependent matches are compared using the wind data as is.

The projection of non-Aeolus winds $y$ onto the Aeolus HLOS direction $x_{dir}^{HLOS}$ is provided in Eqs. (A1-4):

$$u_y = -y_{spd} \sin(y_{dir}) \quad (A1)$$

$$v_y = -y_{spd} \cos(y_{dir}) \quad (A2)$$

$$y_{spd}^{HLOS} = (-u_y \sin(x_{dir}^{HLOS})) + (-v_y \cos(x_{dir}^{HLOS})) \quad (A3)$$

$$y_{dir}^{HLOS} = x_{dir}^{HLOS} \quad (A4)$$

where $y_{spd}$ and $y_{dir}$ are the non-Aeolus wind speed and direction, respectively, $u_y$ and $v_y$ are the u- and v-components of the non-Aeolus wind, $y_{spd}^{HLOS}$ is the non-Aeolus wind vector projected onto the HLOS direction, and $y_{dir}^{HLOS}$ is the new direction of the non-Aeolus wind and is equal to the HLOS direction $x_{dir}^{HLOS}$.